\documentclass[oneside,12pt]{article}
\usepackage{amsthm}
\usepackage{a4}
\usepackage{amssymb}
\usepackage[dvips]{color}
\usepackage{graphicx}

\def\frac#1#2{%
  \mathchoice{%
    {\textstyle\mathstrut#1\over\textstyle\mathstrut#2}}%
    {{\textstyle\mathstrut#1\over\textstyle\mathstrut#2}}%
    {{\scriptstyle\mathstrut#1\over\scriptstyle\mathstrut#2}}%
    {{\mathstrut#1\over\mathstrut#2}%
  }%
}

\newcommand{\im}{{\rm Im}}
\newcommand{\re}{{\rm Re}}
\newcommand{\Var}{{\rm Var}}

\begin{document}
\begin{center}
{\Large\bf Spectral Density of Sample Covariance Matrices of Colored Noise}
\end{center}
\medskip

\begin{center}
{\noindent\bf Emil Dolezal$^{1}$, Petr Seba$^{2,3}$}
\smallskip

{
\small\it
\noindent $^{1}$Faculty of Nuclear Sciences and Physical Engineering, Czech Technical University, Prague, Czech Republic\\
\noindent $^{2}$Department of Physics and Informatics, University of Hradec Kralove, Hradec Kralove, Czech Republic\\
\noindent $^{3}$Institute of Physics, Academy of Sciences of the Czech Republic, Prague, Czech Republic\\
}
\end{center}

\begin{abstract}
We study the dependence of the spectral density of the covariance
matrix ensemble on the power spectrum of the underlying multivariate
signal. The white noise signal leads to the celebrated
Marchenko-Pastur formula. We demonstrate results for some colored
noise signals.
\end{abstract}

\begin{center}
\sc 1. Introduction
\end{center}
The covariance matrix is a fundamental object in the multivariate
statistics and probability theory. A sample covariance matrix use
only part of the data and is determined by the number of samples.
But it has the same population size as the covariance matrix. When
the population size is not large and the number of sampling points
is sufficient the sample covariance matrix is a good approximate of
the covariance matrix. Unfortunately, we usually investigate data
with a sampling rate that is not sufficient and select the number of
samples to be comparable with the population size. In this case the
sample covariance matrix is no longer a good approximation to the
covariance matrix.

Marchenko and Pastur \cite{MarPas} were discussing  a limiting case
when the ratio $p$ between the population size $m$ and the number of
samples $n$ remains constant and $n$ grows without bounds. They
studied the sample covariance matrix $c$ defined by the formula

\begin{equation}\label{defcovmat}
c_{i,j}=\sum_{k=1}^{n}x^{i}_{k}x^{j}_{k}\ \ \ \ {\rm or}\ \ \ \ c=xx^{T}
\end{equation}
where $x^{i}_{k}$ stands for the normalized (i. e. with zero-mean)
independent and identically distributed random data. The upper and
lower indices denote the population and sample index respectively.

The spectral density of $c$ depends in the limit only on the
variance $\sigma^{2}$ of $x$ and on the population-to-sample ratio
$p$

\begin{equation}\label{MarPasDens}
\rho(\lambda)=\left \{
\begin{array}{cc}
\frac{1}{2\pi\lambda p\sigma^{2}}\sqrt{(b-\lambda)(\lambda-a)} & a\leq \lambda\leq b\\ \\
0 & \lambda<a\ \bigvee\ \lambda>b
\end{array}
\right .,
\end{equation}
where $a=\sigma^{2}\left( 1-\sqrt{p} \right )^{2}$,
$b=\sigma^{2}\left( 1+\sqrt{p} \right )^{2}$. For $p>1$, there is an
additional Dirac measure at $\lambda=0$ of mass $1-\frac{1}{p}$.

The formula (\ref{MarPasDens}) describes the spectral density of the
sample covariance matrices of a white noise signal. So the power
spectrum of the signal vector $x^{i}_{k}$  is constant. In many
situations however the signal is not accessible directly. What is
actually measured is its filtered image. For instance if we deal
with the EEG signal we do not measure directly the cerebral signal
but only its image filtered through the tissues in the skull. The
natural question is of course to what degree the spectral density of
the sample covariance matrix depends on such signal filtering. We
show that spectral density (\ref{MarPasDens}) is universal in
certain circumstances and that it represents a special case of the
general probability distribution which depends on the power spectrum
of the signal.

\begin{center}
\large\sc 2. Signal frequency analysis and the covariance matrix
spectral density
\end{center}
The measuring device has a finite sampling rate that leads to a
discrete set of the measured values. For that reason we will use a
discrete Fourier transformation (DFT) for the frequency analysis. In
our notation DFT is defined as

\begin{equation}\label{DFT}
X_k=\frac{1}{\sqrt{n}}\sum_{j=1}^{n}x_j e^{-i\omega_k t_j},\ \omega_k=(k-1)\frac{2\pi f}{n},\ t_j=\frac{j-1}{f},
\end{equation}
where $f$ is the sampling rate.

The Fourier transform $X$ of a real vector $x$ is  complex and
fulfills

\begin{equation}\label{DFTprop}
X_i=\overline{X_{n-i}},\ 1<i\leq n.
\end{equation}

For real $x$ it is therefore useful to use another transformation

\begin{equation}\label{DFTNew}
\widetilde{X}_i=\sqrt{2}\cdot\left \{
\begin{array}{cc}
\re(X_i) & 1< i\leq\left [ \frac{n}{2}\right ]\\ \\
\im\left (X_{n-i}\right ) & \left [
\frac{n}{2}\right ]+1< i \leq n
\end{array}
\right .,
\end{equation}
where $[a]$ means the integer part of $a$. The remaining two
elements are defined separately. Since $X_{1}$ is real and equal to
the sum of $x_{i}$, we define $\widetilde{X}_{1}=X_{1}$. For even
$n$  we take $\widetilde{X}_{\frac{n}{2}+1}=X_{\frac{n}{2}+1}$ since
 $X_{\frac{n}{2}+1}$ is real. For odd $n$ we use
$\widetilde{X}_{\frac{n+1}{2}}~=~\sqrt{2}\re\left
(X_{\frac{n+1}{2}}\right )$.

The transformed vector $\widetilde{X}$ is real and contains the full
information on the frequency properties of the original vector $x$.
The definition of the covariance matrix (\ref{defcovmat}) can be
easily rewritten using the discrete Fourier transformation and the
transformation (\ref{DFTNew}):

\begin{equation}\label{defcovmatmod}
c_{i,j}=\sum_{k=1}^{n}x^{i}_{k}x^{j}_{k}=\sum_{k=1}^{n}X^{i}_{k}\overline{X^{j}_{k}}=\sum_{k=1}^{n}\widetilde{X}^{i}_{k}\widetilde{X}^{j}_{k}\ \ {\rm or}\ \ c=xx^{T}=XX^{*}=\widetilde{X}\widetilde{X}^{T},
\end{equation}
where the rows of the matrices $X$ and $\widetilde{X}$ are the transformed rows of the matrix $x$.

Colored noise is a random signal  with a non-flat power spectrum. We
are interested in the question how the profile of the power spectrum
influence the spectral density of the sample covariance matrix. In
what follows we assume that the data matrix $x$ has independent rows
with identical power spectra and zero mean. Then the elements of the
matrix $\widetilde{X}$ are also of mean zero - see the definition
(\ref{defcovmatmod}). Moreover the elements in the rows of the
matrix $x$ are independent. The transform  $\widetilde{X}$ leads
therefore also to a matrix with independent rows. Since the signal
phase is random we get

\begin{equation}\label{VarRelCond}
\left <\re\left (X_{k}^{i}\right  )^2\right >=\left <\im\left
(X_{k}^{i}\right )^2\right >
\end{equation}
and hence

\begin{equation}\label{VarRel}
\left <\left (\widetilde{X}_k^{i}\right )^{2}\right >=\left <\left (\widetilde{X}_{n-k}^{i}\right )^{2}\right >, \ \ \left [ \frac{n}{2}\right ]+1< k \leq n,\ \ 1 \leq i \leq m,
\end{equation}
where the angle brackets denote the sample mean. To find the
spectral density of the covariance matrix ensemble we use now the a
theorem  of Girko \cite{Girko}.

\bigskip
{ {\noindent\bf Theorem 1:} Let $A$ be a $m\times [cm]$ random
matrix with independent entries of a zero-mean that satisfy the
condition

\begin{equation}\label{GirkoVarCond}
m\Var(A_{ij})<B,
\end{equation}
for some bound $B<\infty$. Moreover, let for each $m$ be $v_m$ a
function
$v_m{:}\,\left [0,1\right ]\times \left [0,c\right
]\rightarrow \mathbb{R}$ defined by:

\begin{equation}\label{GirkoDefV}
v_m(\mu,\nu)=m\Var(A_{ij}),\ \ \ \ \frac{i}{m}\leq \mu\leq\frac{i+1}{m},\ \frac{j}{m}\leq \nu\leq \frac{j+1}{m}
\end{equation}
and suppose that $v_m$ converges uniformly to a limiting bounded function $v$ for $m\to\infty$.
Then the limiting eigenvalue distribution $\rho(\lambda)$ of the covariation matrix $AA^{\rm T}$ exists and for every $\tau\geq 0$ satisfies:

\begin{equation}\label{GirkoEq1}
\int\limits_{0}^{\infty}\frac{\rho (\lambda){\rm d}\lambda}{1+\tau\lambda}=\int\limits_{0}^{1}u(\mu,\tau){\rm d}\mu,
\end{equation}
with $u(\mu,\tau)$ solving the equation
\begin{equation}\label{GirkoEq2}
u(\mu,\tau)=\frac{1}{1+\tau\int\limits_{0}^{c}{\frac{v(\mu,\nu){\rm d}\nu}{1+\tau\int\limits_{0}^{1}u(\xi,\tau)v(\xi,\nu){\rm d}\xi}}}.
\end{equation}
The solution of the equation (\ref{GirkoEq2}) exists and is unique
in the class of functions $u(\mu,\tau)\geq 0$, analytical on $\tau$ and
continuous on $\mu\in \left [0,1\right ]$. }
\bigskip

Let us use this theorem taking  $A=\widetilde{X}$. We immediately
see that $c=\frac{n}{m}$. Since all the rows of the matrix
$\widetilde{X}$ have  identical power spectra, the function
$v(\mu,\nu)$ will not depend on $\mu$. The equation (\ref{GirkoEq2})
shows that the function $u(\mu,\tau)$ is also $\mu$ independent.
Inserting $v(\nu)$ and $u(\tau)$ into the equations (\ref{GirkoEq1})
and (\ref{GirkoEq2}), we get

\begin{equation}\label{GirkoEq11}
\int\limits_{0}^{\infty}\frac{\rho (\lambda){\rm d}\lambda}{1+\tau\lambda}=u(\tau)
\end{equation}
and

\begin{equation}\label{GirkoEq22}
u(\tau)=\frac{1}{1+\tau\int\limits_{0}^{c}{\frac{v(\nu){\rm d}\nu}{1+\tau u(\tau)v(\nu)}}}.
\end{equation}

The spectral density is determined  by the function $v(\nu)$ (that
itself is a function of the power spectrum).
However, to solve the  equations (\ref{GirkoEq11}) and
(\ref{GirkoEq22}) for a general power spectrum profile is extremely
difficult. So in next chapters we will try to get an exact formula
for the spectral density at least in the simplest cases.

\begin{center}
\large\sc 3. Generalized white noise
\end{center}

\begin{figure}[!ht]
\begin{center}
\includegraphics[width=14.4cm,height=5.4675cm]{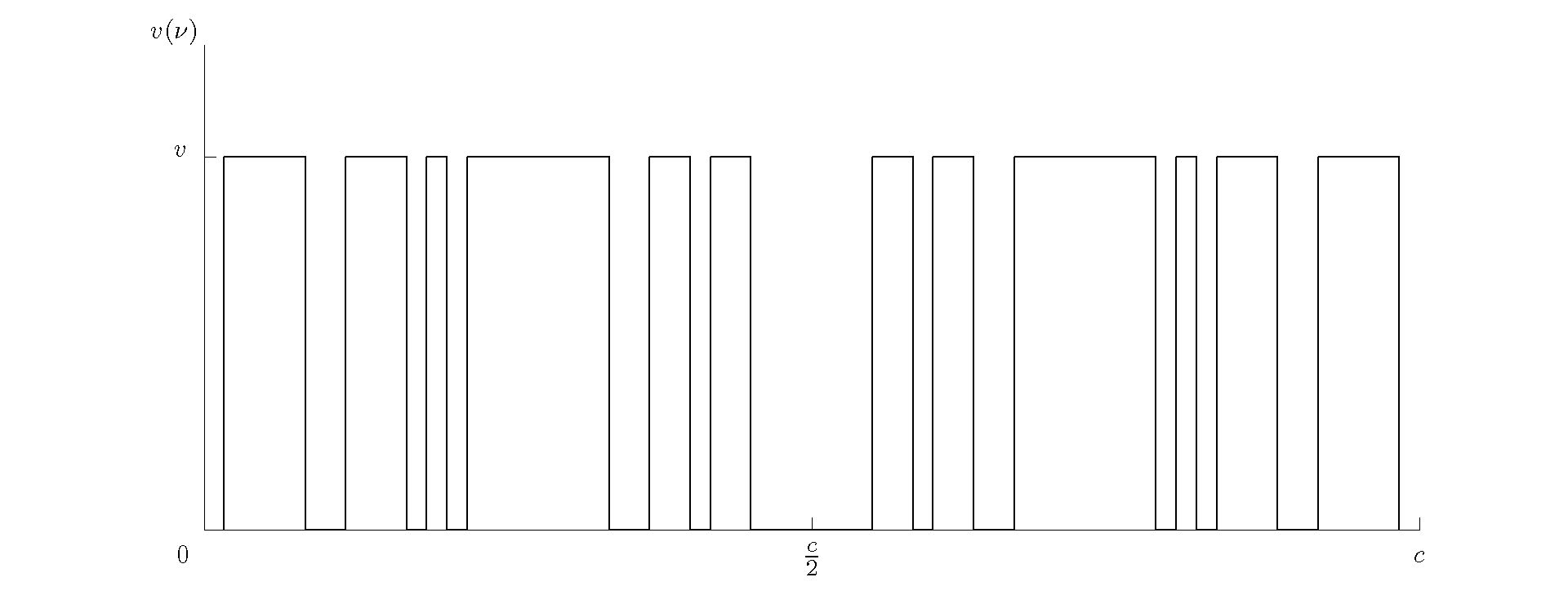}
\caption{\leftskip=1cm\rightskip=1cm An example of the function
$v(\nu)$  for the generalized white noise.}\label{ZobecBilSum}
\end{center}
\end{figure}

Consider a situation when the signals from several sources come into
one given point. Every sources produce a noise in a specific
frequency bands and the frequency bands are disjoint. Further, the
intensity of all sources is the same. The total incoming signal has
gaps in the power spectrum. The function $v(\nu)$ (see the definition
in the Theorem 1.) is a step function with steps of an equal hight,
see the figure (\ref{ZobecBilSum}).

In order to evaluate the spectral density we have to know the size
$d$ of the support of $v(\nu)$ (i.e. the sum of the lengths of all
intervals where $v(\nu)$ is nonzero) and the value $v$ of the
function $v(\nu)$ on this support (the function $v(\nu)$ is constant
on the support). The solution  of the equation (\ref{GirkoEq22})
gives
\begin{equation}\label{GenWhiteNoiseU}
u(\tau)=\frac{-\left [1+v\tau(d-1)\right ]+ \sqrt{\left [1+v\tau(d-1)\right
]^{2}+4v\tau}}{2v\tau}
\end{equation}
and the integral equation (\ref{GirkoEq11}) can be transformed into
the form
\begin{equation}\label{GirkoEq111}
\int\limits_{0}^{\infty}\frac{\rho (\lambda){\rm d}\lambda}{\nu+\lambda}=\frac{u\left (\frac{1}{\nu}\right )}{\nu}=\widetilde{u}(\nu),
\end{equation}

The generalized Stieltjes transform
\begin{equation}\label{Stielt}
G(\xi)=\int\limits_{0}^{\infty}\frac{F(\nu){\rm d}\nu}{(\nu+\xi)^{q}},\ \ \left | \arg \xi \right |<\pi,
\end{equation}
has an inverse \cite{Schwarz}
\begin{equation}\label{InvStielt}
F(\nu)=-\frac{1}{2\pi i}\nu^{q}\int\limits_{\mathcal{C}}\frac{G^{\prime}(yw){\rm d}w}{(1+w)^{1-q}},
\end{equation}
where $q>0$ and  $\mathcal{C}$ is a contour starting at the point
$w=-1$ and encircling the origin in the counterclockwise sense. For
$q=1$ the eq. (\ref{InvStielt}) can be explicitly evaluated:

\begin{equation}\label{InvStieltSpec}
F(\nu)=\lim_{\varepsilon\rightarrow 0^{+}}\frac{1}{2\pi i}\left ( G(-\nu-i\varepsilon)-G(-\nu+i\varepsilon) \right )
\end{equation}
for $\nu>0$.

We find the spectral density $\rho(\lambda)$ as an inverse Stieltjes
transform with $q=1$. The limit in (\ref{InvStieltSpec}) with
 $G=\widetilde{u}$ and $F=\rho$ leads to

\begin{equation}\label{HustZobec}
\rho(\lambda)=\left \{
\begin{array}{cc}
\frac{1}{2\pi\lambda v}\sqrt{(\lambda_1-\lambda)(\lambda-\lambda_2)} & \lambda_1\leq \lambda\leq \lambda_2\\ \\
0 & \lambda<\lambda_1\ \bigvee\ \lambda>\lambda_2
\end{array}
\right .,
\end{equation}
where $\lambda_{1,2}=v\left (1\mp\sqrt{d}\right )^{2}$. For $d<1$ ,
there is an additional Dirac measure at $\lambda=0$ of mass $1-d$.

The above formula is exactly equal to the Marchenko-Pastur result
(\ref{MarPasDens}). The existence of the Dirac measure is
consequence of singularity of covariance matrices for $d<1$.

\begin{figure}[!ht]
\begin{center}
\includegraphics[width=14.4cm,height=9.612cm]{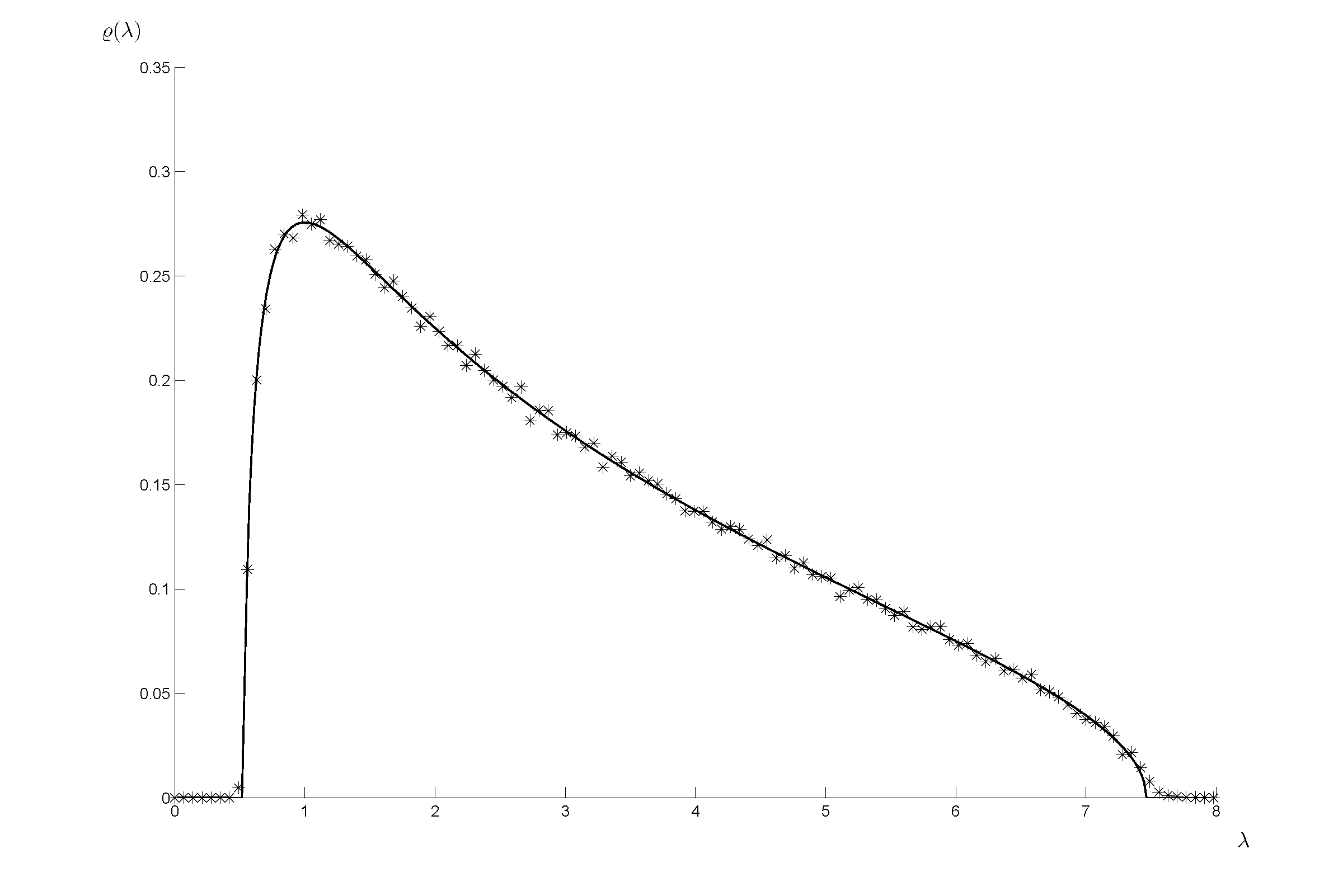}
\caption[E]{\leftskip=1cm\rightskip=1cm Spectral density. Numerical
results (stars) are compared with the theoretical results for
$\rho(\lambda)$ with the parameters $v=1$ and $d=3$.}
\label{GrafDistZobBilSum}
\end{center}
\end{figure}

The interesting point is that while the Marchenko-Pastur result was
derived for a white noise signal (i.e. the power spectrum was
constant over the whole frequency range) we get  the same result
also when the power spectrum  has a finite support and contains a
finite number of gaps. Moreover - the exact position of the gaps is
irrelevant and the result depends on the total support size only.

\begin{center}
\large\sc
4. Colored noise
\end{center}

Let us now pass to the case when the highs of the power spectrum
segments are unequal. This is a quite general case since in fact any
power spectrum profile can be approximated by a step function.

Inserting the function $v(\nu)$ into the integral (\ref{GirkoEq22}) and
using the definition (\ref{GirkoEq111}) gives

\begin{equation}\label{SpecBarU}
\tau\widetilde{u}(\tau)+\sum_{i=1}^{K}d_i-1=\sum_{i=1}^{K}\frac{d_i}{1+v_i\widetilde{u}(\tau)},
\end{equation}
where $K$ is number of the nonzero segments in the function
$v(\nu)$. In what follows the symbols  $v$ and $d$ denote vectors
(in contrast to their previous meaning as constants) with elements
$v_i$ and $d_i$ denoting the highs and lengths of segments
respectively.

To solve the equation (\ref{SpecBarU}) means to find the roots of a
polynomial of degree $(K+1)$.  This cannot be done explicitly.
However - in similarity to the previous case with the bars of equal
hight -   the solution of the equation (\ref{SpecBarU}) does not
depend on the exact location of power spectrum bars and is positive
on the positive real axis.

As an illustration we give the formula for the case with two steps:

\begin{equation}\label{SpecSpektrDistrHod}
K=2,\ v=\left (1,\frac{1}{2}\right ),\ d=\left (\frac{1}{2},\frac{1}{2}\right ).
\end{equation}

The spectral density is than

{
\footnotesize
\begin{equation}\label{SpecSpektrDistr}
\rho(\lambda)=\sqrt {{\frac { \left( -\sqrt [3]{-27\,\lambda-8\,{\lambda}^{3}+36\,{\lambda}^{2}+27+3\,\sqrt {3}\sqrt {\lambda}\sqrt {54-16\,{\lambda}^{3}+72\,{\lambda}^{2}-81\,\lambda}}+2\,\lambda-3 \right) ^{2}}{8\pi^2\lambda\sqrt [3]{-27\,\lambda-8\,{\lambda}^{3}+36\,{\lambda}^{2}+27+3\,\sqrt {3}\sqrt {\lambda}\sqrt {54-16\,{\lambda}^{3}+72\,{\lambda}^{2}-81\,\lambda}}}}}.
\end{equation}
}

\begin{figure}[!ht]
\begin{center}
\includegraphics[width=14.4cm,height=9.612cm]{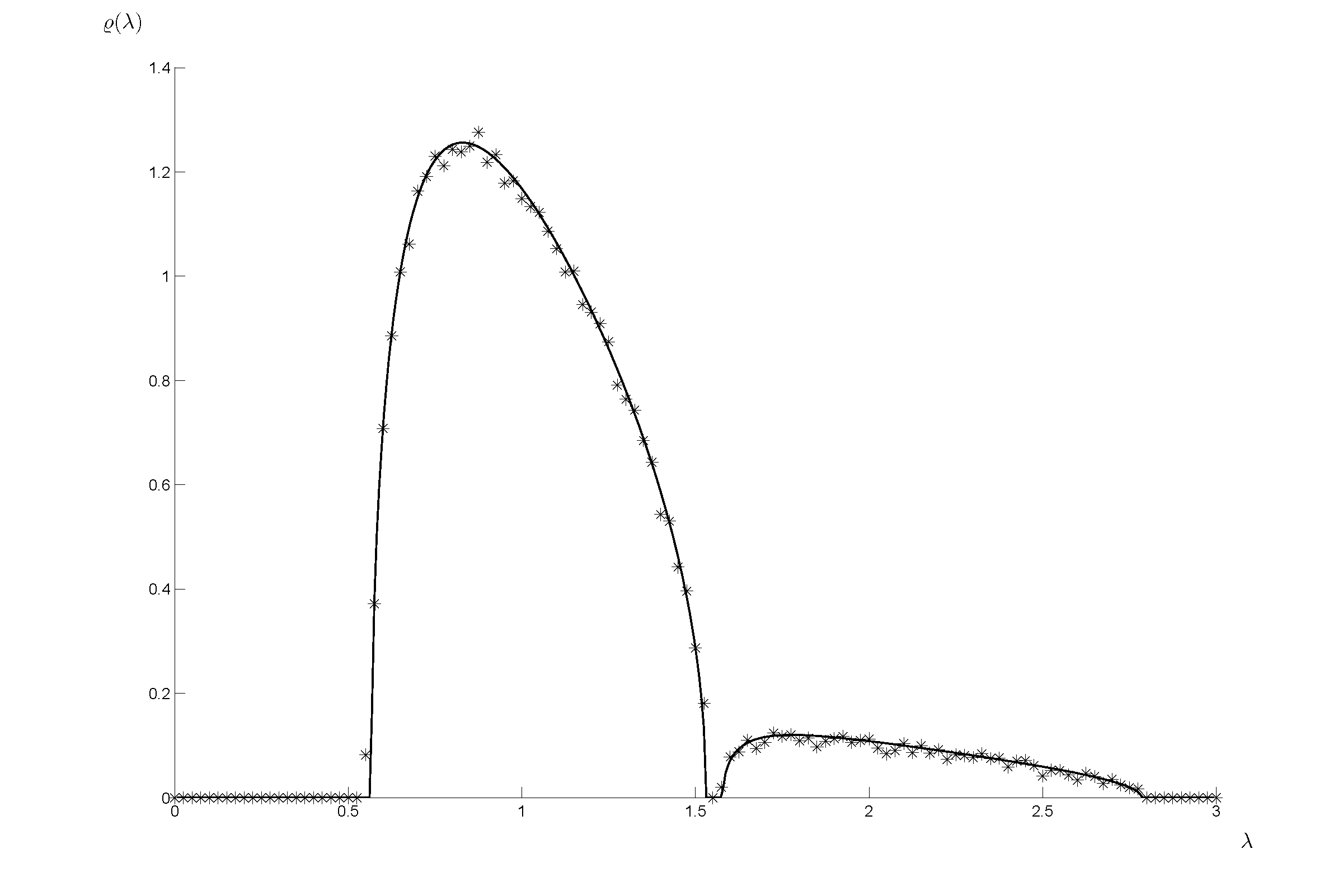}
\caption[a]{\leftskip=1cm\rightskip=1cm  Spectral density. The
numerical results (stars) are compared with the theoretical
prediction of $\rho(\lambda)$ for $K=2$, $v=\left (1,\frac{1}{15}\right )$,
$d=\left(\frac{1}{10},15\right )$.}
\label{GrafSpecBarSumB}
\end{center}
\end{figure}

The spectral density of the covariance matrices will in this case
again not depend on the exact location of power spectrum steps.
Also the order of the steps is not important.  Moreover  segments of
the same hight can be linked into one segment with the width equal
to the sum of the widths of the two particulars segments. In this
sense the spectral density does not depend on the reshuffling of the
power spectrum.

\begin{center}
\large\sc 5. Summary
\end{center}

The spectral density of the covariance matrix is used in many fields
of physics and economy (see \cite{BurdaBurda}, \cite{Burda},
\cite{Dumitriu}). To analyze the system the power spectrum of the
signal has to be taken into account.  An example is the spectral
analysis of the EEG signal \cite{duge},\cite{Seba}.

The power spectrum directly influence the signal correlation
properties. For instance the particular matrix elements of the
covariance matrix depend on it. Nevertheless the spectral density of
the covariance matrix ensemble remains nearly invariant.

In the presented paper we discuss  the spectral density of the
covariance matrix and its dependence on the power spectrum profile
of the underlying signal. The results show that the spectral density
is invariant under the reshuffling of its power spectrum
coefficients and hence independent on the exact spectral profile of
the signal.

\begin{center}
\large\sc 6. Acknowledgement
\end{center}

The research was supported by  the Ministry of Education, Youth and 
Sports within the project LC06002.

\end{document}